\documentclass[12pt]{iopart}
\usepackage{amssymb}
\usepackage{blindtext}
\usepackage{booktabs}
\usepackage{cite}
\usepackage{datetime2}
\usepackage{hyperref}
\usepackage{lineno}
\hypersetup{
    colorlinks=true, %set true if you want colored links
    linktoc=all,     %set to all if you want both sections and subsections linked
    linkcolor=black,  %choose some color if you want links to stand out
    citecolor=blue,
    filecolor=blue,
    urlcolor=blue
}
\usepackage[utf8]{inputenc}
\usepackage{iopams}
\usepackage{setstack}
\usepackage{titlesec}
\usepackage{xcolor}
\usepackage{xfrac}
\usepackage{xspace}
\usepackage{upgreek}
\usepackage{verbatim}

\date{\today}
\newcommand{\px}{\ensuremath{p_{x}}\xspace}
\newcommand{\py}{\ensuremath{p_{y}}\xspace}
\newcommand{\pz}{\ensuremath{p_{z}}\xspace}
\newcommand{\kt}{\ensuremath{k_\mathrm{T}}\xspace}
\newcommand{\PT}{\ensuremath{p_\mathrm{T}}\xspace}
\newcommand{\gauss}{\ensuremath{\mathcal{N}(0,1)}\xspace}
\newcommand{\dkl}{\ensuremath{D_\mathrm{KL}}\xspace}
\newcommand{\emd}{EMD\ensuremath{_{\textrm{\footnotesize sum}}}\xspace}
\newcommand{\unit}[1]{\ensuremath{\text{\,#1}}\xspace}
\newcommand{\GeV}{\ensuremath{\,\text{Ge\hspace{-.08em}V}}\xspace}
\newcommand{\TeV}{\ensuremath{\,\text{Te\hspace{-.08em}V}}\xspace}

\begin{document}

\title[LHC Hadronic Jet Generation Using ConVAE+NFs]{LHC Hadronic Jet Generation Using Convolutional Variational Autoencoders with Normalizing Flows}
\author{Breno Orzari$^1$, Nadezda Chernyavskaya$^2$, Raphael Cobe$^1$, Javier Duarte$^3$, Jefferson Fialho$^1$, Dimitrios Gunopulos$^4$, Raghav Kansal$^3$, Maurizio Pierini$^2$, Thiago Tomei$^1$, Mary Touranakou$^{2,4}$}
\address{$^1$Universidade Estadual Paulista, S\~{a}o Paulo/SP - CEP 01049-010, Brazil \\
$^2$European Organization for Nuclear Research (CERN), CH-1211 Geneva 23, Switzerland \\
$^3$University of California San Diego, La Jolla, CA 92093, USA\\
$^4$Department of Informatics and Telecommunications, National and Kapodistrian University of Athens, Athens 157 72, Greece}
\ead{breno.orzari@cern.ch}
\vspace{10pt}
\begin{indented}
\item[]June 2023
\end{indented}

% \linenumbers

\section*{Abstract}

In high energy physics, one of the most important processes for collider data analysis is the comparison of collected and simulated data. 
Nowadays the state-of-the-art for data generation is in the form of Monte Carlo (MC) generators. 
However, because of the upcoming high-luminosity upgrade of the LHC, there will not be enough computational power or time to match the amount of needed simulated data using MC methods. 
%An approach that is under study is the use machine learning generative methods to fulfill this task instead. 
An alternative approach under study is the usage of machine learning generative methods to fulfill that task.
Since the most common final-state objects of high-energy proton collisions are hadronic jets, which are collections of particles collimated in a given region of space, this work aims to develop a convolutional variational autoencoder (ConVAE) for the generation of particle-based LHC hadronic jets. 
Given the ConVAE's limitations, a normalizing flow (NF) network is coupled to it in a two-step training process, which shows improvements on the results for the generated jets. 
The ConVAE+NF network is capable of generating a jet in $18.30 \pm 0.04\unit{$\upmu$s}$, making it one of the fastest methods for this task up to now.

% \newpage

% \tableofcontents

\newpage

\section{Introduction}
In the CERN LHC~\cite{LHC1,LHC4}, approximately one billion proton-proton ($pp$) collisions occur every second.
Those collisions generate outgoing particles that are measured by dedicated, custom-made particle detectors.
The main data stream collected by such a detector is close to 1.4\unit{kHz}.
After the upcoming LHC upgrade to the High-Luminosity LHC (HL-LHC)~\cite{HL-LHC} this number will increase by a factor of 3.5--5. 
In high energy physics (HEP), an important part of the data analysis is the simulation of $pp$ collisions that are used to test analysis techniques, estimate the particle detectors' efficiencies and resolutions, and make comparisons with collected data.
Usually, the number of simulated events is an order of magnitude higher than the number of recorded events in real data, such that the statistical uncertainty of the simulated events doesn't dominate the overall uncertainty. For simulating collision data, Monte Carlo (MC) event generators are used. However, the increased granularity of the detectors
and the higher complexity of the collision events after the HL-LHC upgrade pose significant challenges that cannot be solved by increasing the computing resources alone~\cite{Software:2815292}.

One of the most common final-state objects in high-energy $pp$ collisions is an energetic cluster of hadrons, known as a hadronic jet~\cite{Jet1}.  
Jets can be described by the features of their constituent particles, mainly their four-momentum.
The particle-flow (PF)~\cite{PFlow} algorithm is used to combine information obtained by several sub-detectors to precisely measure each particle's properties. 
Jet clustering algorithms~\cite{Jetography} can be used to cluster the particle constituents into a jet and obtain its relevant properties for physics analysis such as its invariant mass, transverse momentum (\PT), energy, azimuthal angle $\phi$, and pseudorapidity $\eta$\footnote{ 
In the LHC experiments, the origin of the right-handed coordinate system is the center of the local $pp$ collision region, the $y$ axis is defined vertically upward, and the $z$ axis is along the proton beam direction.
Additional coordinates are the azimuthal angle $\phi$ and pseudorapidity $\eta = -\ln \tan (\theta/2)$, where $\theta$ is the polar angle.
}.

Being able to accelerate the process of jet simulation in a particle detector would be of great advantage to mitigate the needed overhead in computing resources.
One of the avenues that shows promising results is the use of generative machine learning (ML) algorithms to perform the jet simulation. The goal of this program is to maintain the accuracy of MC methods while making the process orders of magnitude faster. 
Several approaches using neural networks (NNs) have been studied and applied for this purpose, using generative adversarial networks (GANs)~\cite{MPGAN,EpicGAN}, normalising flows~\cite{Kach:2022uzq}, and score-based diffusion models~\cite{Leigh:2023toe,Mikuni:2023dvk}.
Another promising technique is the use of variational autoencoders (VAEs)~\cite{VAESparse}, which have yet to demonstrate high fidelity results.

In this work, a new method for the generation of hadronic jets from noise is presented. It is based on a VAE coupled with a normalizing flow (NF) and trained in a two-step process. The paper is organized as follows. Section~\ref{section:related} contains a literature overview of ML applied to HEP, mainly focusing on generative methods for jets simulation. In Section~\ref{section:dataset} the dataset and the model are described. Section~\ref{section:lossfunc} details the custom loss function and the evaluation metrics; results are presented and discussed in Section \ref{section:results}. Finally in Section~\ref{section:summary}, we summarize the work and give an outlook for using the new method.

\section{Related Work}
\label{section:related}

ML techniques have been applied to hadronic jets generation following several distinct approaches. 
Some of the first attempts~\cite{LAGAN,CaloGAN,3DGAN,BIBAE,VAE} were performed using convolutional neural networks (CNNs) because of their impressive results in computer vision and the fact that an image-like representation of a jet arises naturally from the particle detectors' data. 
Other jet representations such as vectors of energy deposits in calorimeters or high-level jet characteristics have also been studied~\cite{ATLAS,LHCb}.

Recently, particle-based jet datasets have also been extensively used by the ML community in the context of jets generation~\cite{MPGAN,EpicGAN,FastVAE,VAESparse}.
They are composed of the jets constituent particles' properties and give a detailed description of particles distribution inside the jet, which is known as the jet substructure~\cite{JetSub1,JetSub2}.
GANs based on graph networks~\cite{MPGAN,EpicGAN} have provided inspiring results when aiming to generate particle-based jets, retaining great similarity between generated jets and MC simulated jets while improving the speed of simulation by 4 to 5 orders of magnitude depending on the number of particles inside the generated jet.
Some VAEs based on convolutional networks have also been implemented using the same dataset for jets fast simulation~\cite{FastVAE} and for the jets generation as well~\cite{VAESparse}. 
However, for the latter, there is great room for improvement when comparing the VAE generated jets with MC simulated jets.

The premise of this work is that the VAE by itself is not enough to capture the full characteristics of a hadronic jet, mainly due to the \textit{a priori} choice of the latent space distribution.
On the other hand, a combined VAE+NF approach, which has already been applied for image generation~\cite{VAEFlowIm}, time series data prediction~\cite{VAEFlowPr} and calorimeter shower simulation~\cite{AE+NF1,AE+NF2}, among others, has not yet been applied in the context of hadronic jets generation.
In this work, this combined approach is used to improve the generative capacity of the VAE.

\section{Dataset and Model}
\label{section:dataset}

We use the gluon HLS4ML LHC Jet dataset~\cite{JetData, JediNet} that is composed of $\sim$177k gluon jets, containing the jet constituent particles' three-momenta $(\px, \py, \pz)$. 
These were obtained from simulated $pp$ collisions with a center of mass energy of $\sqrt{s} = 13\TeV$ that produced jets with $\PT \approx 1\TeV$. 
A simplified particle detector parameterization, similar to the apparatuses deployed by the large LHC experiments, is used to simulate particles' interactions with the detector material.
The hadronic jets are reconstructed using the anti-\kt algorithm~\cite{FastJet, Antikt} with a distance parameter $R = 0.8$. 
The 177k samples are divided into $\sim$70\% for training and $\sim$15\% each for validation and testing. 
Even though there is no intrinsic ordering to the particles inside a jet, the particles are ordered in a list by decreasing \PT and only the first 30 particles are used as input to the ML algorithm. 
If a jet contains less then 30 particles, the rest of the list was filled with zeros (zero-padding). A feature-wise standardization was performed to bound the values of the three-momentum to be in the range [0.0, 1.0] as input to the network.

\subsection{Convolutional VAE}

To accommodate this representation of the input dataset, the chosen model is a VAE~\cite{AEvarBayes,TutorialVAE,IntroVAE} composed of convolution layers (ConVAE). 
The VAE is composed of three components: an encoder, which learns to compress the representation of the input data into a lower dimensional representation, a latent space, which stores this lower dimensional representation into values that closely follow a probability distribution defined by the user, and a decoder, which decodes the latent space values to the output. 
The probability distribution of the latent space values has to be differentiable, easy to implement, and easy to sample from.
The most common choice is the standard Gaussian \gauss. 
%The goal of the VAE model is to learn the representation of input data into the latent space and how to decode this compressed information to go back to an output that is as similar as possible to the input
The goal of the VAE is to produce output data whose distribution is as similar as possible to that of the input data.
In addition, it is desirable to be able to generate new data samples resembling the training data by sampling from \gauss and feeding those values into the decoder.

The network was implemented in the PyTorch library~\cite{Pytorch}. 
In this work, the encoder and decoder structures are mirrored, and only the encoder will be described. 
It is composed of a given number of consecutive two-dimensional convolution layers where the last convolution layer is flattened and introduced into two linear layers, where the latter is input into the latent space.
An activation function is applied after each layer, except for the last linear layer of the encoder.
The input and output layers of the network have a fixed size of 3$\times$30.
The activation function in between each layer, number of convolutional layers in the encoder and decoder, kernel size, latent dimension size, number of filters, and number of nodes in the linear layers between encoder (decoder) and the latent space are hyperparameters to be optimized.
The activation function of the last deconvolution layer of the decoder was chosen to be the sigmoid~\cite{Sigmoid} function, to bound the output values in between 0.0 and 1.0, given the feature-wise standardization.

In the encoder architecture, the number of input nodes of the first linear layer is fixed by the size of the last convolutional layer, while the number of output nodes of the second linear layer is fixed as two times the number of dimensions of the latent space to provide the means and standard deviations of the latent distributions\footnote{A.k.a., the reparametrization trick~\cite{IntroVAE}.}.
In the decoder architecture, the number of input nodes in the first linear layer is fixed by the latent space size, and the number of output nodes in the second linear layer is fixed by the size of the first deconvolution layer.
The stride and padding of the convolution layers were kept as 1 and 0, respectively.
The first convolutional layer has 1 input filter and $N_\mathrm{filters}$ output filters, and the rest of the convolutional layers have an input number of filters equal to 2$^{\ell-1} \times N_\mathrm{filters}$ and an output number of filters equal to 2$^{\ell} N_\mathrm{filters}$, where $\ell$ is the layer number.

Hyperparameter optimization was performed using the Optuna package~\cite{Optuna}, which allows for fast convergence thanks to aggressive pruning based on intermediate results and easy parallelization. 
Each ConVAE was set to go through 300 trials of the optimization, where each trial was executed for 300 epochs of the training. 
At training time, the batch size was kept fixed at 100 samples and the Adam optimizer~\cite{Adam} was used with a learning rate (LR) that was set by the hyperparameter optimization.
During the optimization, the evaluation metric (described in Section~\ref{section:lossfunc}) was minimized using the value evaluated on the validation dataset after every 5 epochs. 
The ConVAE was retrained with the best set of hyperparameters for 1500 epochs, and the best model was chosen as the one that exhibited the smallest value of the metric.

\section{ConVAE Loss Function and Evaluation Metric}
\label{section:lossfunc}

The VAE loss function is given by the following expression~\cite{AEvarBayes}:
\begin{equation}
L(\theta,\phi) = - \mathbb{E}_{z \sim q_{\phi}(\mathbf{z}|\mathbf{x})} \left[ \log \ p_{\theta}(\mathbf{x}|\mathbf{z}) \right] + \dkl(q_{\phi}(\mathbf{z}|\mathbf{x})\parallel p(\mathbf{z})),
\end{equation}
where $\mathbf{x}$ is the vector of data values, $\mathbf{z}$ is the vector of latent space values, $q_{\phi}(\mathbf{z}|\mathbf{x})$ the variational posterior (i.e., the encoder), $p_\theta(\mathbf{x}|\mathbf{z})$ the conditional likelihood (i.e., the decoder), and $p(\mathbf{z})$ the prior distribution.
The first term on the right-hand side can be seen as the reconstruction loss $L_{\mathrm{rec}}$, measuring the distance between the output produced by the network and the input data.
The second term \dkl is the Kullback-Leibler divergence~\cite{kldiv} that measures the difference between the probability distributions of the encoder output values and the latent space values.
The distribution of the latent space values is defined \textit{a priori}, and the distribution of the encoder output values is one that is easy to work with, but still complex enough to describe its output, where the usual choice is a multivariate Gaussian.

The minimization of this loss function ensures that the output will be as close as possible to the input, while the values of the latent space closely follow the chosen distribution. In addition, one can add a hyperparameter $\beta$~\cite{betaVAE} to control the importance of each term in the loss. The loss function can be written as:
\begin{equation}
    \label{eq:VAELoss}
    L_\mathrm{VAE} = (1-\beta)L_\mathrm{rec}  + \beta \dkl
\end{equation}
The reconstruction loss term was customized to the task of particle generation. The loss function that compares output particles with input dataset particles was chosen to be the Chamfer distance~\cite{Chamfer}, also referred to as nearest neighbor distance (NND). This loss function is preferred over the commonly-used mean-squared error (MSE) function because it is permutation invariant, which is desirable since particles in a jet have
no intrinsic ordering. The NND loss is expressed as:
\begin{equation}
    L^\mathrm{NND} =  { \sum}_{k} \left[ { \sum}_{i \in \mathcal{J}_{k}} \min_{j \in \hat{\mathcal{J}}_{k}} D(\vec{p}_i, \vec{\hat{p}}_j) + { \sum}_{j \in \hat{\mathcal{J}}_{k}} \min_{i \in \mathcal{J}_{k}} D(\vec{p}_i, \vec{\hat{p}}_j) \right],
\end{equation}
where $i$ and $j$ are indices of the particles in the input and output samples, respectively, $k$ is the index of a given jet, $\mathcal{J}_{k}$ represents the $k$-th jet in the dataset, hat distinguishes between input (without) and output (with) objects, and $D(\vec{p}_i, \vec{\hat{p}}_j)$ is the Euclidean distance between input and output particles, treating each as a vector in $(\px, \py, \pz)$ space. The first term finds the closest output particle to a given input particle, the second term finds the closest input particle to a given output particle, and their distances are summed.

The loss $L^\mathrm{NND}$ alone was not enough to provide good quality of the generated jets. Therefore, physics-inspired constraints were included in the form of distances between the input and output jet properties. 
The best results were achieved using a combination of the jet mass and \PT terms using MSE as a distance function. The additional loss term was added as follows 
\begin{equation}
    L^\mathrm{J} = { \sum}_{k} \left[ \gamma_{\PT}\mathrm{MSE}(p_{
\mathrm{T}_{k}}, \hat{p}_{\mathrm{T}_{k}}) + \gamma_{m}\mathrm{MSE}(m_{k}, \hat{m}_{k}) \right].
\end{equation}
where the parameters $\gamma_{\PT}$ and $\gamma_m$ were used to weight each contribution.
Finally, the combined reconstruction loss is given by
\begin{equation}
    L_\mathrm{rec} = \alpha L^\mathrm{NND} + \gamma L^\mathrm{J},
\end{equation}
in which $\alpha$ and $\gamma$ were used to weight the importance of each contribution to the loss function. 
The full set of loss parameters $(\alpha, \beta, \gamma, \gamma_{\PT}, \gamma_m)$ was optimized with Optuna.

The hyperparameter optimization procedure was carried out in two steps.
In the first, only the loss parameters and the optimizer learning rate were optimized while the network parameters were fixed at some reasonable choices established after manual tests.
In the second, all the network parameters were optimized given the best set of loss parameters. 
This was done to ensure convergence of the optimization procedure in a feasible amount of time given the amount of resources available.

The evaluation metric chosen to quantitatively measure the generation capabilities of the distinct models was the earth mover's distance (EMD)~\cite{emd}, or 1-Wasserstein distance, calculated for each histogram of the input and output jets mass, energy, \PT, $\eta$, and $\phi$, and summed to get the \emd.
The choice of the variables was based on relevant quantities for jet-based searches in LHC experiments~\cite{JetSearch1,JetSearch2}.
This metric was used as the hyperparameter optimization objective, the method to choose the best model, and to compare different approaches.

\subsection{Normalizing Flows}

When the ConVAE is optimized for reconstruction, i.e., the value of $\beta$ in Equation~(\ref{eq:VAELoss}) is small, a looser constraint is applied to ensure the latent space values follow a given probability distribution.
In this case, the probability distribution of those values, $p(\mathbf{z})$, favours optimizing the reconstruction rather than strictly matching the imposed prior.
These reconstruction-imposed latent space distributions are generally unknown and one cannot easily sample from them.

A normalizing flow~\cite{NF1,NF2} is a transformation of a simple probability distribution (e.g., \gauss) into a more complex distribution (and vice versa) through a sequence of invertible and differentiable mappings usually learned by training a NN.
Therefore, we can combine the ConVAE approach with a NF to find the transformation from the more complex latent space distribution to a simpler one, such as $p(\mathbf{z}) \overset{f}\longrightarrow  u(\mathbf{x})$ where $u(\mathbf{x}) = \gauss$.
Since these transformations are invertible by construction, it is possible to start from values that follow the simple distribution, pass it through the inverse of the transformation and obtain values that follow the complex and unknown distribution as $u(\mathbf{x}) \overset{{f^{-1}}}\longrightarrow p(\mathbf{z})$.
The network is trained by maximizing the right-hand side of Equation~(\ref{eq:NFLoss}).
\begin{equation}
\label{eq:NFLoss}
    \ln(p(\mathbf{z})) = \ln(u(f(\mathbf{z}))) + \ln \left| \text{det} \left( \frac{\partial f(\mathbf{z})}{\partial \mathbf{z}} \right) \right|,
\end{equation}
where $\partial f(\mathbf{z})/\partial \mathbf{z}$ is the Jacobian of the transformation.

The NF network we use is based on the RealNVP network\footnote{The implementation of the RealNVP network in this work was performed following the GitHub repository \url{https://github.com/ispamm/realnvp-demo-pytorch}.}~\cite{RealNVP1,RealNVP2} that makes use of simple analytic expressions for each intermediate transformation
\begin{equation}
\label{eq:NFcoup1}
    y_{1:d} = x_{1:d},
\end{equation}
\begin{equation}
\label{eq:NFcoup2}
    y_{d+1:D} = x_{d+1:D} \cdot \text{exp}(s(x_{1:d})) + t(x_{1:d}),
\end{equation}
using the coupling layers~\cite{coupling}, where only a subset of the input data undergoes the transformations while the rest remains unchanged, and the subsets are permuted every epoch to ensure that all input data goes through the flows.
The parameters $s$ and $t$ are obtained as the outputs of two multilayer perceptrons (MLPs), which can be as complex as needed.

\begin{figure}[ht]
    \centering
    \includegraphics[width=1.0\textwidth]{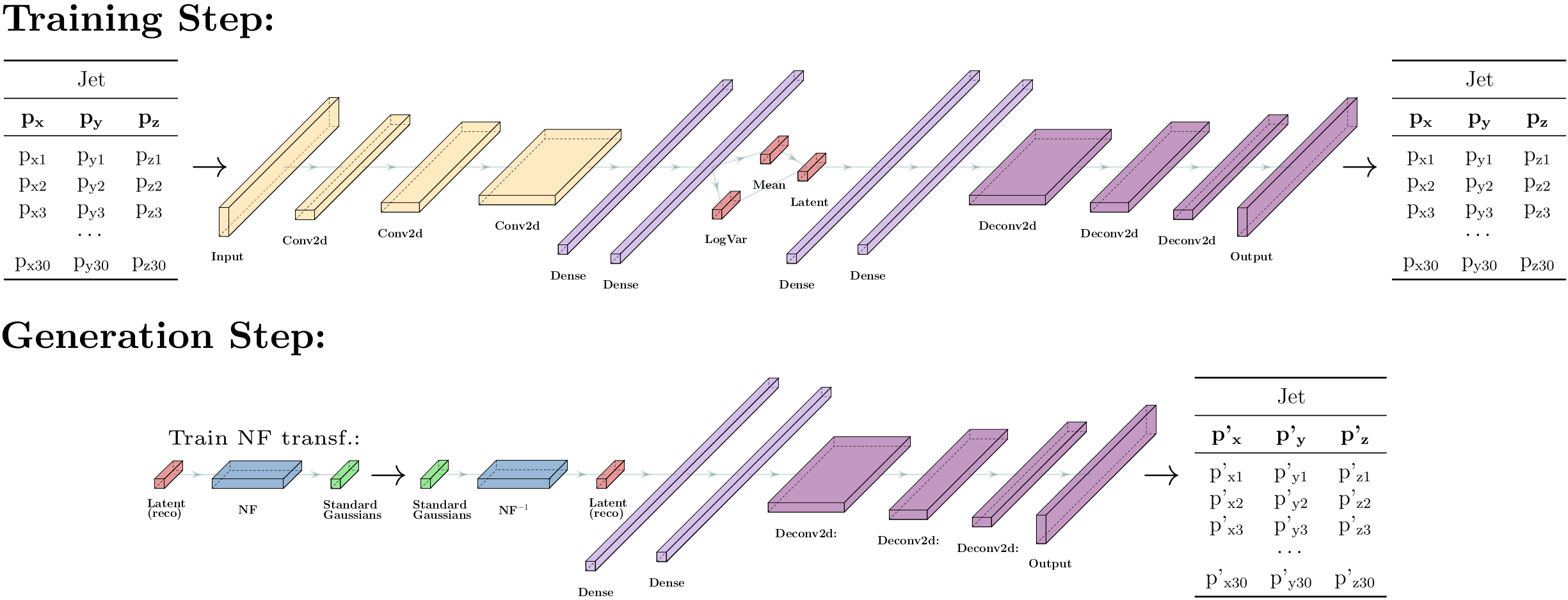}
    \caption{Illustration of ConVAE+NF network training scheme.}
    \label{fig:trainsch}
\end{figure}

To combine the ConVAE with the NF network (ConVAE+NF), a two-step training was performed. First, the hyperparameters of the ConVAE were optimized and the resulting best model was trained for 1500 epochs. The model with the smallest \emd when comparing the input with reconstructed jets in the validation dataset, calculated every 5 epochs, was saved. 
The latent space values of the saved network when receiving as input the training and validation jet datasets were extracted to generate new training and validation datasets for the NF.
A diagram containing the full training procedure is displayed in Figure~\ref{fig:trainsch}.

The NF network was implemented using the PyTorch library.
The Adam optimizer was chosen to update the network parameters by minimizing the negative of the mean of the expression in Equation~(\ref{eq:NFLoss}) as the loss function.
The learning rate, number of flows and the number of layers and nodes of the two MLPs that determine $s$ and $t$ were hyperparameters to be optimized.
The ReLU~\cite{ReLU} activation function was used for each layer of the MLPs, and for $s$ the hyperbolic tangent was used after its last layer to constrain its values to be in the range $(-1.0, 1.0)$.
The number of input and output nodes of the $s$ and $t$ networks were set to match the number of latent dimensions of the ConVAE.

The metric used to evaluate the combination of both networks as the hyperparameter optimization objective was \emd.
After every four epochs, $N_\mathrm{latent}$ values were sampled from \gauss and input into the inverse of the NF network, whose output was given as input to the decoder of the ConVAE tuned for reconstruction, and its output was compared with jets from the validation dataset.
The optimization was run for 300 trials with 20 epochs of NF training per trial.
The best set of hyperparameters was used to train the final NF network for 100 epochs, and the model that exhibited the smallest metric value, checked after every epoch, was saved.

\section{Results and Discussion}
\label{section:results}

\subsection{Hyperparameter optimisation}

As described above, the hyperparameter optimization was performed using the Optuna framework and executed in steps to speed up the convergence of the method given the limited computational resources. The ranges considered for each hyperparameter during the optimization for both types of NN are shown below.
In the ConVAE case, the set of parameters to be optimized was: the learning rate of the optimizer (Adam), the loss function parameters $\alpha$, $\beta$, $\gamma$, $\gamma_{\PT}$, and $\gamma_m$, and the architecture parameters as the number of latent dimensions, number of filters, kernel size related to the number of particles to convolve, number of nodes in linear layers, number of convolution layers, and activation function.
For the 2D kernel size $(k_1, k_2)$, only $k_2$, the number of particles to convolve, was varied, while $k_1$, the number of particle features to convolve, was set to 3 for the first (last) convolution (deconvolution) layer of the encoder (decoder) and 1 otherwise.

The ranges of the parameters $\gamma_{\PT}$ and $\gamma_m$ were determined by the dataset because jet \PT and mass have different scales, which imply different magnitudes for those loss components.
The range for $\gamma_{\PT}$ was set from 20\% smaller to 20\% larger than unity, while the range for $\gamma_m$ was set from 20\% smaller to 20\% larger than the ratio of the mean of the jet \PT and mass.
For the loss function hyperparameter optimization, the network architecture hyperparameters were fixed at the reasonable choices displayed in Table~\ref{tab:ConVAECommon}.
For both ConVAE networks (ConVAE and ConVAE+NF approaches) almost all the ranges of the hyperparameters for the optimization were the same,
and are shown in Table~\ref{tab:commonRanges}. %:
\begin{comment}
\begin{itemize}
    \item LR: [10$^{-5}$, 10$^{-1}$] in log scale;
    \item $\alpha$: [0.1, 1.0];
    \item $\gamma$: [0.1, 1.0];
    \item $\gamma_{\PT}$: [0.8, 1.2];
    \item $\gamma_{m}$: [9.986, 14.98];
    \item Latent space dimension: [10, 300] in steps of 10;
    \item Number of filters: [5, 100] in steps of 5;
    \item Kernel size: [1, 8] in steps of 1;
    \item Number of linear nodes: [100, 3000] in steps of 100;
    \item Number of convolution layers: [1, 4] in steps of 1;
    \item Activation function: one of ReLU, GeLU, LeakyReLU, SELU, or ELU.
\end{itemize}
\end{comment}
The only exception was for the loss hyperparameter $\beta$ that, due to the difference in the magnitudes of $L_{\mathrm{rec}}$ and \dkl, needed to be much closer to 1.0 for the ConVAE tuned for generation, where the optimization range was [0.9, 1.0] in log scale, and much closer to 0.0 for the ConVAE tuned for reconstruction, where the optimization range was [0.0, 0.1] in log scale.

\begin{table}[ht]
    \centering
    \addtolength{\tabcolsep}{-0.2em}
    \begin{tabular}{c c c c c c}
    \toprule
         $N_\mathrm{latent}$ & Filters & Kernel & $N_\mathrm{linear}$ & Layers & Act. Func.  \\ \midrule
          30 & 50 & 5 & 1500 & 3 & ReLU \\
          \bottomrule
         %& 
    \end{tabular}
    \caption{Set of common architecture hyperparameters for ConVAE loss parameters optimization.}
    \label{tab:ConVAECommon}
\end{table}

\begin{table}[ht]
    \centering
    \addtolength{\tabcolsep}{-0.2em}
    \begin{tabular}{c c c c c }
    \toprule
         LR & $\alpha$ & $\gamma$ & $\gamma_{\PT}$ & $\gamma_m$ \\ \midrule
         10$^{-5}$--10$^{-1}$ & 0.1--1.0 & 0.1--1.0 & 0.8--1.2 & 9.986--14.98 \\ \bottomrule\toprule
         $N_\mathrm{latent}$ & Filters & Kernel & $N_\mathrm{linear}$ & Layers  \\ \midrule
          10--300 & 5-100 & 1--8 & 100--3000 & 1--4 \\
    \bottomrule
         %& 
    \end{tabular}
    \caption{Ranges of common architecture hyperparameters for ConVAE loss parameters optimization.
    The choice of activation function was also optimized, chosen from one of
    ReLU, GeLU, LeakyReLU, SELU or ELU.}
    \label{tab:commonRanges}
\end{table}

Since the number of hyperparameters of the RealNVP network is much smaller than for the ConVAE, its hyperparameter optimization was performed only in one step.
The set of hyperparameters that were optimized was: the learning rate of the optimizer (Adam); and the network architecture hyperparameters as the number of flows, number of linear layers for the $s$ and $t$ MLPs and the number of nodes in each layer of those MLPs.
For both the RealNVP network used together with the ConVAE and the one used by itself, the ranges of the hyperparameter optimization are shown in Table~\ref{tab:NFRanges}. %:
\begin{table}[ht]
    \centering
    \begin{tabular}{c c c c}
    \toprule
         LR & $N_{\mathrm{flows}}$ & Layers & $N_\mathrm{linear}$  \\ \midrule
          10$^{-5}$--10$^{-1}$ & 5--100 & 1--4 & 50--400 \\
          \bottomrule
         %& 
    \end{tabular}
    \caption{Ranges of architecture hyperparameters for the RealNVP optimization. 
    The number of nodes in each linear layer ($N_\mathrm{linear}$) is optimized separately for each layer.}
    \label{tab:NFRanges}
\end{table}
\begin{comment}
\begin{itemize}
    \item LR: [10$^{-5}$, 10$^{-1}$] in log scale;
    \item Number of flows: [5, 100] in multiples of 5;
    \item Number of linear layers of $s$ and $t$ MLPs: [1, 4] in steps of 1;
    \item Number of nodes in each linear layer: [50, 400] in steps of 50 (optimized for each layer).
\end{itemize}
\end{comment}

\subsection{Experiments}

\begin{figure}[ht]
    \centering
    \includegraphics[width=0.29\textwidth]{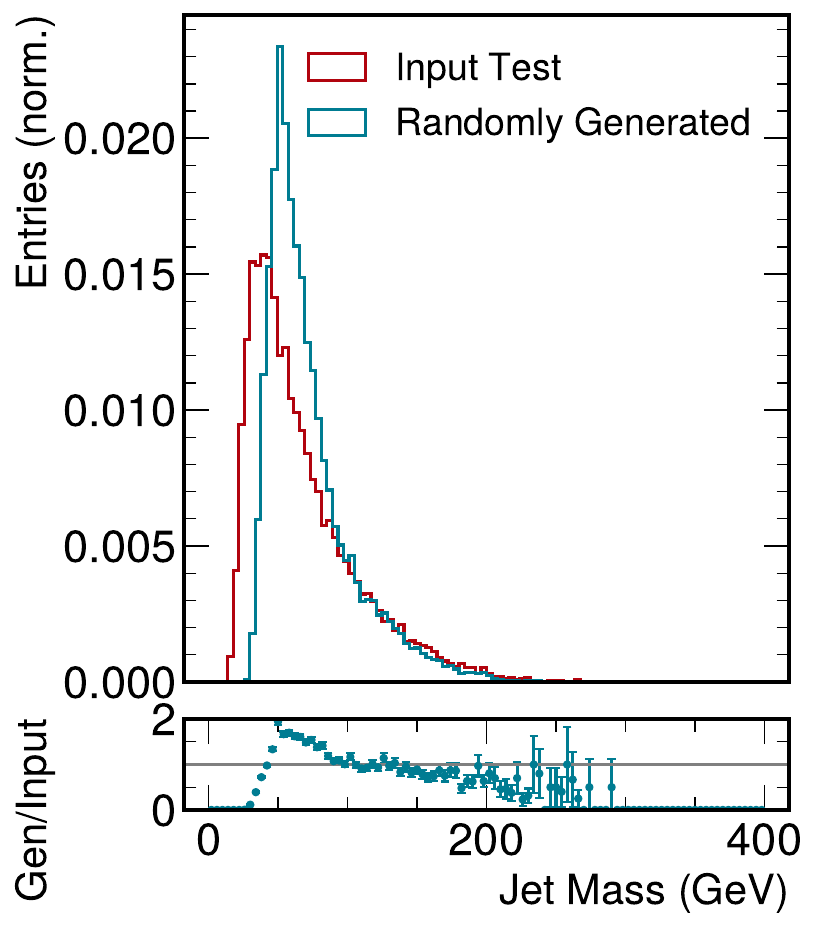}
    \includegraphics[width=0.3\textwidth]{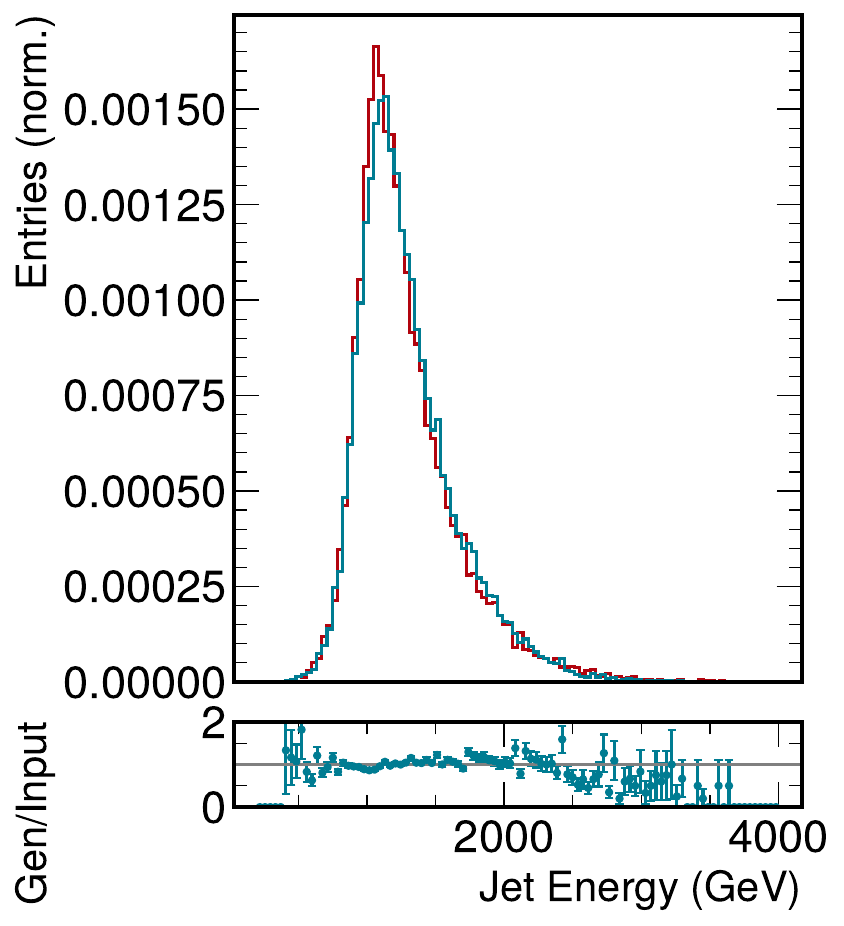}
    \includegraphics[width=0.3\textwidth]{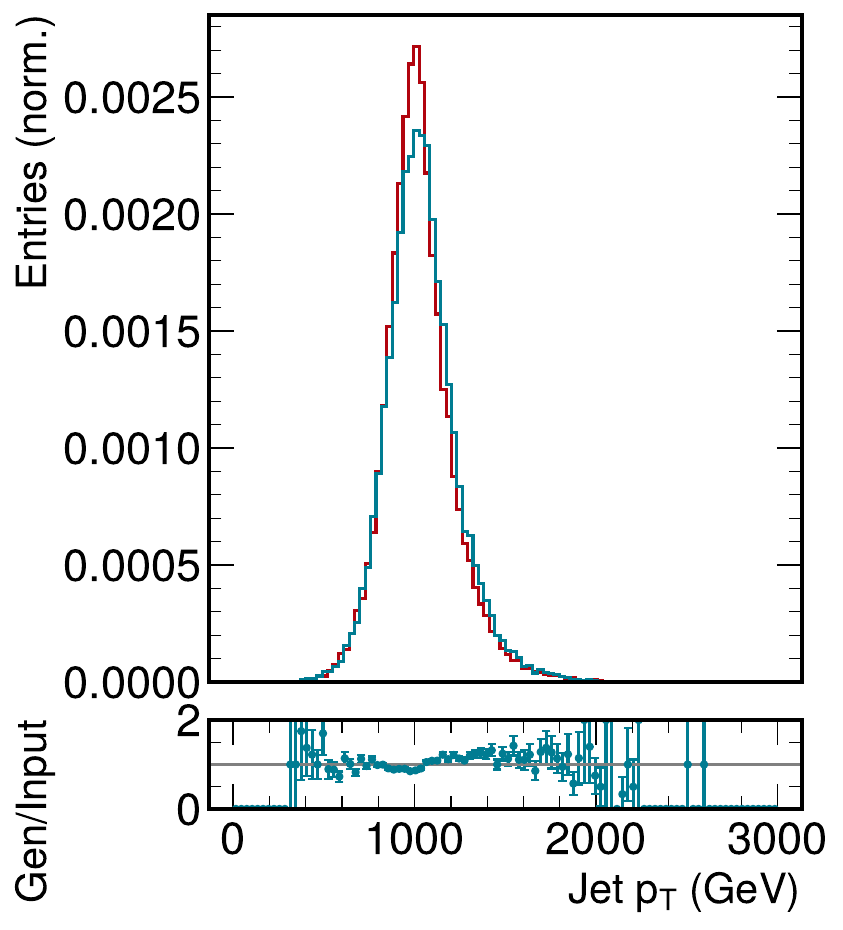} \\
    \includegraphics[width=0.3\textwidth]{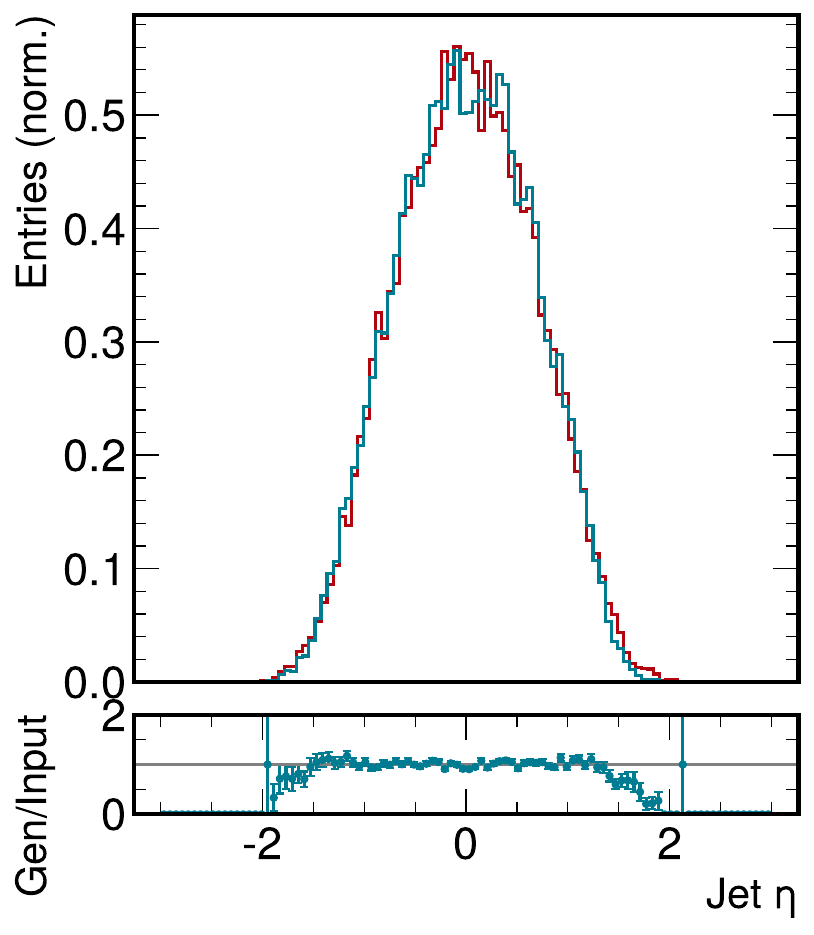}
    \includegraphics[width=0.3\textwidth]{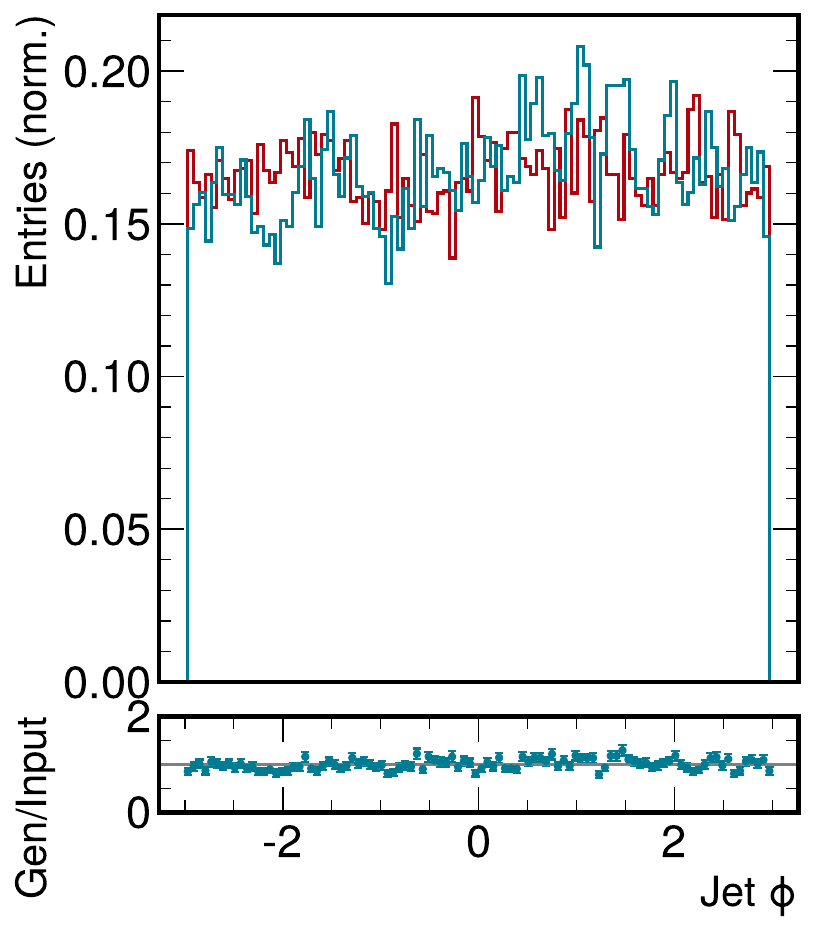}
    \caption{Comparison of input jets variable distributions (red) with randomly generated jets from the ConVAE model (blue). From left to right, top to bottom: mass, energy, \PT, $\eta$, and $\phi$ distributions are displayed. In the subpanels, the ratio $\sfrac{\mathrm{generated}}{\mathrm{input}}$ is shown.}
    \label{fig:ConVAEdist}
\end{figure}

We experiment with three approaches: a baseline ConVAE-only model, the ConVAE+NF model, and another baseline NF-only model.
The final set of optimized hyperparameters for the former is displayed in Table~\ref{tab:ConVAE}, which result in an \emd value of 0.0033.
Figure~\ref{fig:ConVAEdist} shows the comparisons of the distributions of input and output jet mass, \PT, energy, $\eta$ and $\phi$ after hyperparameter optimization. 
Qualitatively, the distributions of jet energy, $\eta$, and $\phi$ are very similar, with differences only at the extremes of the histograms, while for \PT there is a slight difference at the peak of the distribution and small discrepancies at both tails. 
The greatest discrepancies are in the distributions of jet mass, where they are %similar only in the range $80\GeV \lessapprox m_\mathrm{jet} \lessapprox 175\GeV$.
only similar in the approximate range $80\GeV < m_\mathrm{jet} < 175\GeV$.

\begin{table}[ht]
    \centering
    \addtolength{\tabcolsep}{-0.3em}
    \begin{tabular}{c c c c c c c c c c c c}
    \toprule
         LR & $\alpha$ & $\beta$ & $\gamma$ & $\gamma_{\PT}$ & $\gamma_m$ & $N_\mathrm{latent}$ & Filters & Kernel & $N_\mathrm{linear}$ & Layers & Act. Func.  \\ \midrule
          6.1$\times 10^{-4}$ & 0.449 & 0.998 & 0.118 & 0.817 & 11.7 & 190 & 75 & 3 & 1100 & 4 & ReLU \\
          \bottomrule
         %& 
    \end{tabular}
    \caption{Set of optimal hyperparameters for the ConVAE tuned for generation.}
    \label{tab:ConVAE}
\end{table}

\begin{figure}[ht]
    \centering
    \includegraphics[width=0.29\textwidth]{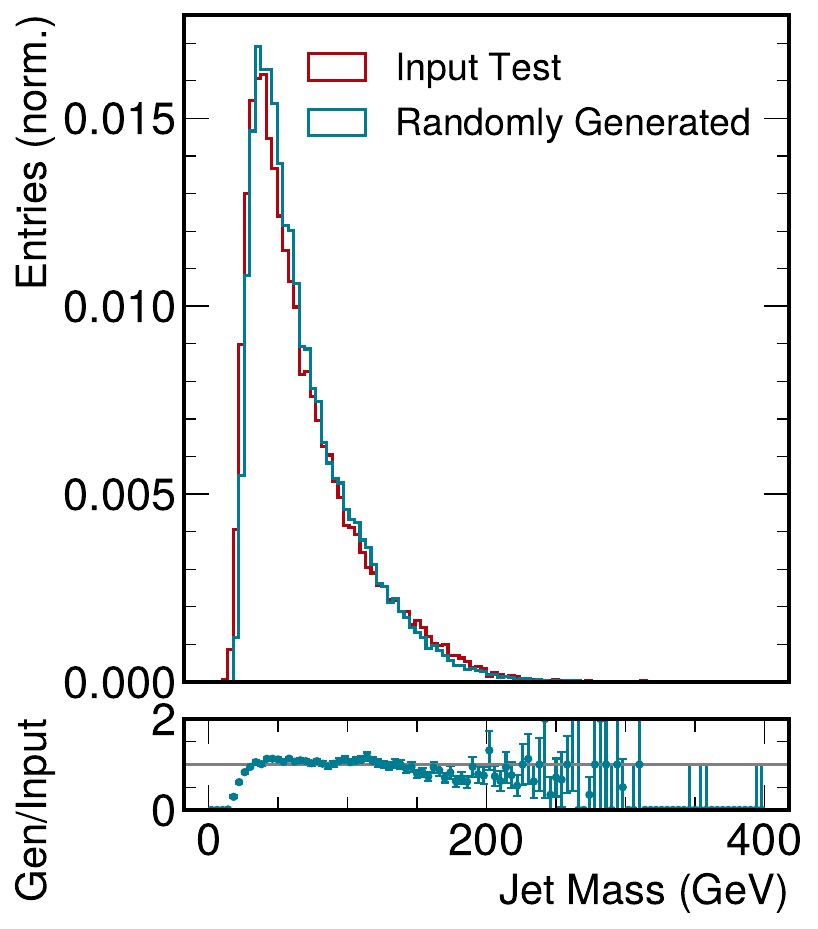}
    \includegraphics[width=0.3\textwidth]{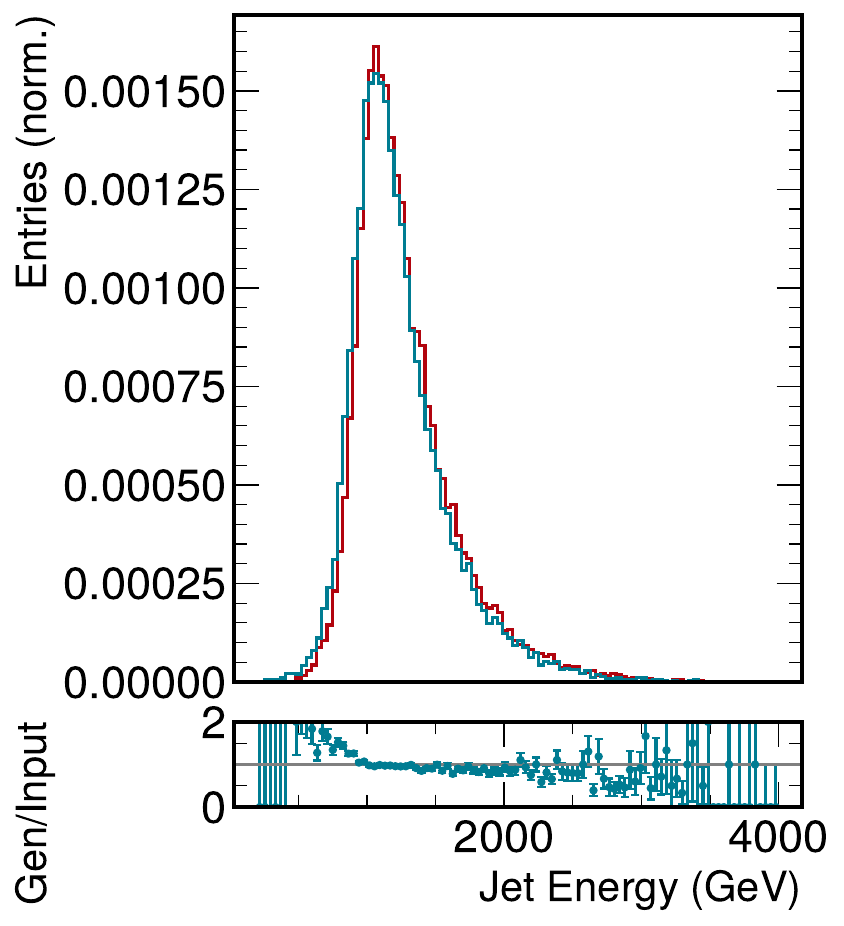}
    \includegraphics[width=0.3\textwidth]{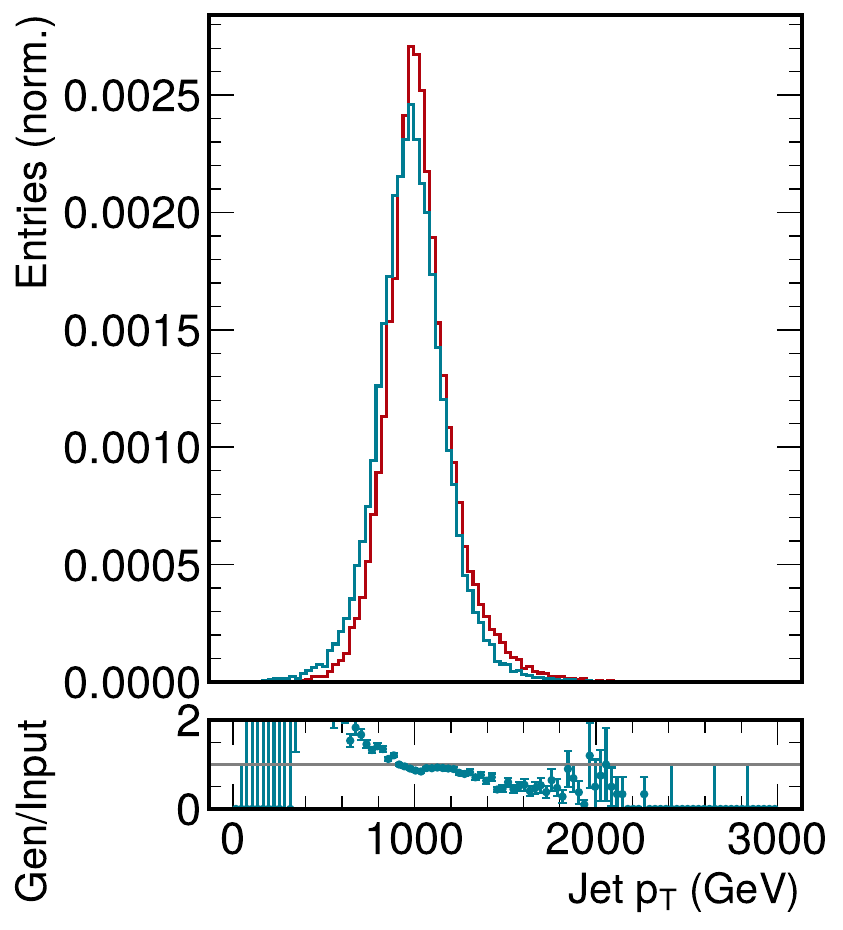} \\
    \includegraphics[width=0.3\textwidth]{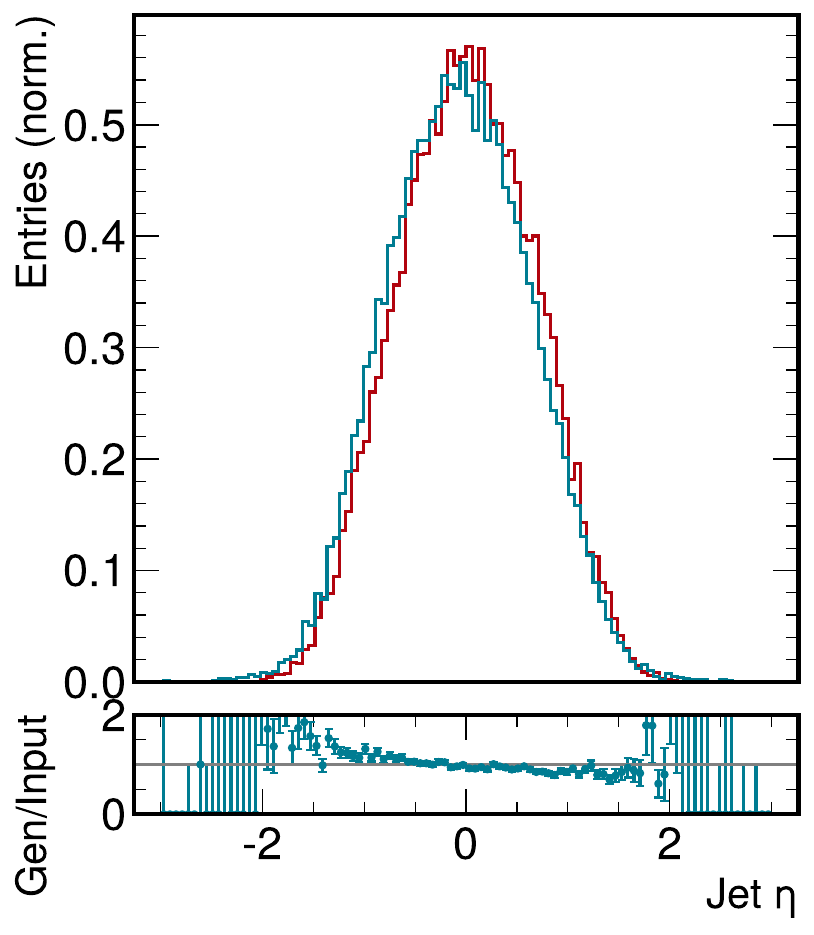}
    \includegraphics[width=0.3\textwidth]{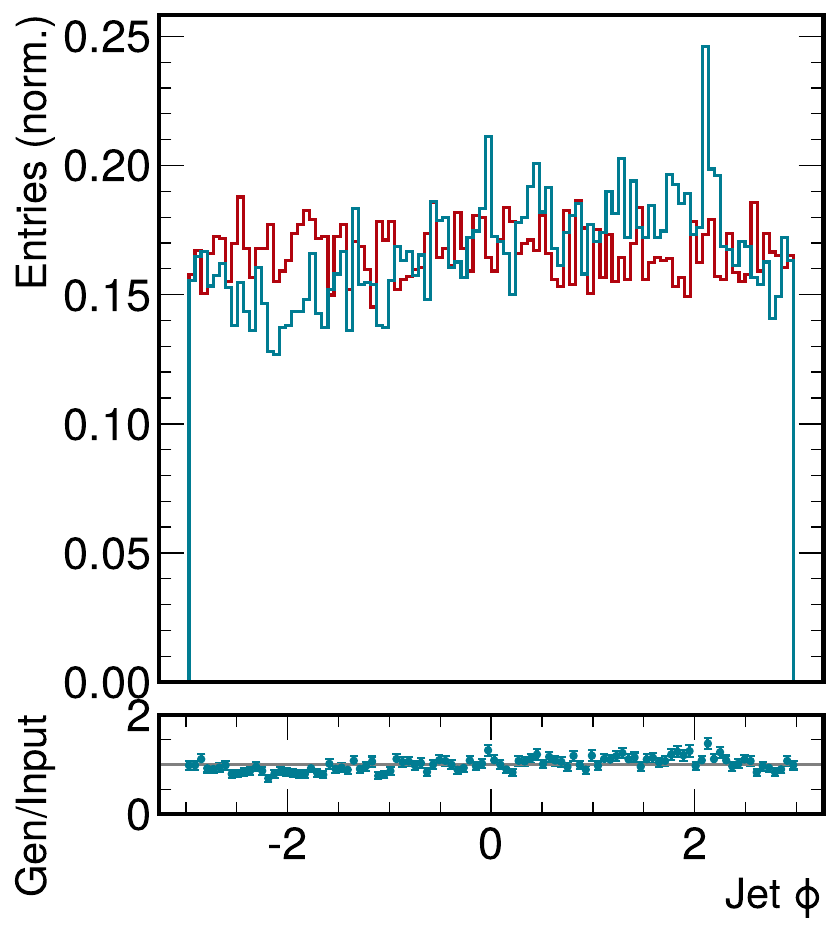}
    \caption{Comparison of input jets variable distributions (red) with randomly generated jets from the ConVAE+NF model (blue). From left to right, top to bottom: mass, energy, \PT, $\eta$, and $\phi$ distributions are displayed. In the subpanels, the ratio $\sfrac{\mathrm{generated}}{\mathrm{input}}$ is shown.}
    \label{fig:ConVAE+NF}
\end{figure}

The complete sets of optimized hyperparameters for the ConVAE+NF model are displayed in Tables~\ref{tab:ConVAE+NF1} and \ref{tab:ConVAE+NF2}, which result in an \emd of 0.0026.
The histograms of the jet variables are displayed in Figure~\ref{fig:ConVAE+NF}.

Based on the \emd, there is a large improvement in the generative capabilities of the ConVAE+NF model compared to the ConVAE model. 
In particular, the agreement in the jet mass distribution improves dramatically, with disagreement only for masses smaller than 25\GeV.
However, there remain notable deviations from unity in the $\eta$, $\phi$, and \PT histograms ratio between randomly generated and input jets.

\begin{table}[ht]
    \centering
    % \addtolength{\tabcolsep}{-0.2em}
    \addtolength{\tabcolsep}{-0.3em}
    \begin{tabular}{c c c c c c c c c c c c}
    \toprule
         LR & $\alpha$ & $\beta$ & $\gamma$ & $\gamma_{\PT}$ & $\gamma_m$ & $N_\mathrm{latent}$ & Filters & Kernel & $N_\mathrm{linear}$ & Layers & Act. Func.  \\ \midrule
          $5.3\times 10^{-4}$ & 0.425 & 0.046 & 0.121 & 1.15 & 11.8 & 180 & 50 & 2 & 3000 & 4 & ReLU \\
          \bottomrule
         %& 
    \end{tabular}
    \caption{Set of optimal hyperparameters for ConVAE tuned for reconstruction.}
    \label{tab:ConVAE+NF1}
\end{table}

\begin{table}[ht]
    \centering
    \begin{tabular}{c c c c}
    \toprule
         LR & $N_{\mathrm{flows}}$ & Layers & $N_\mathrm{linear}$  \\ \midrule
          $2.9\times 10^{-4}$ & 55 & 4 & [400, 350, 400, 400] \\
          \bottomrule
         %& 
    \end{tabular}
    \caption{Set of optimal hyperparameters for the NF to be used on the ConVAE+NF approach.}
    \label{tab:ConVAE+NF2}
\end{table}

\begin{figure}[ht]
    \centering
    \includegraphics[width=0.29\textwidth]{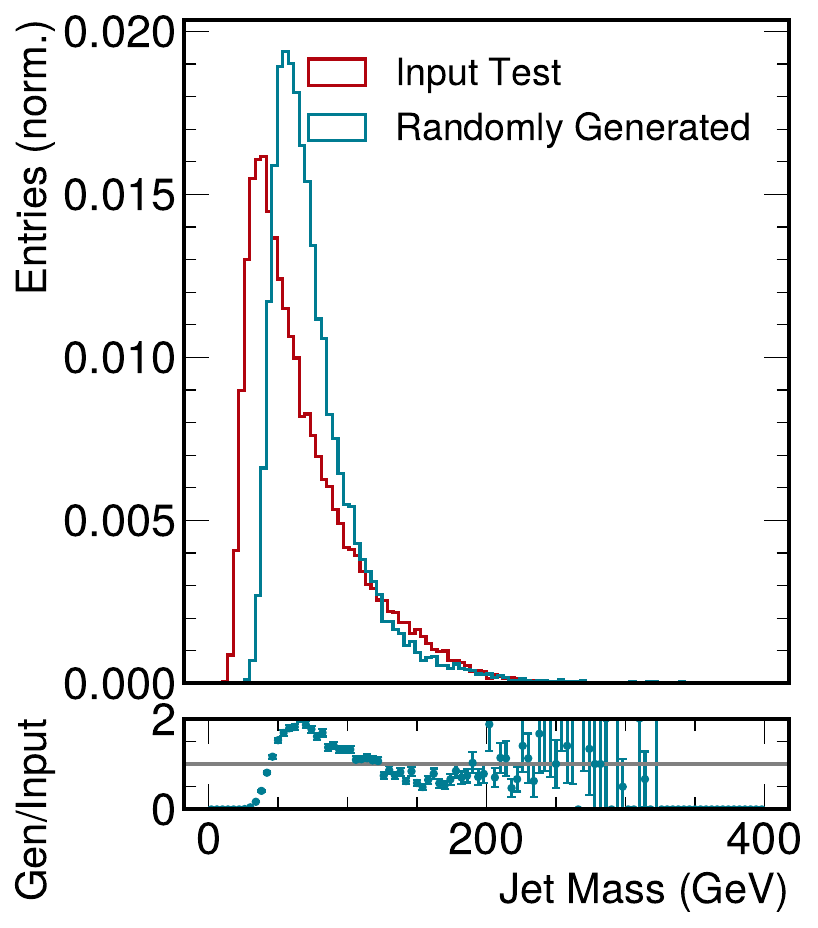}
    \includegraphics[width=0.3\textwidth]{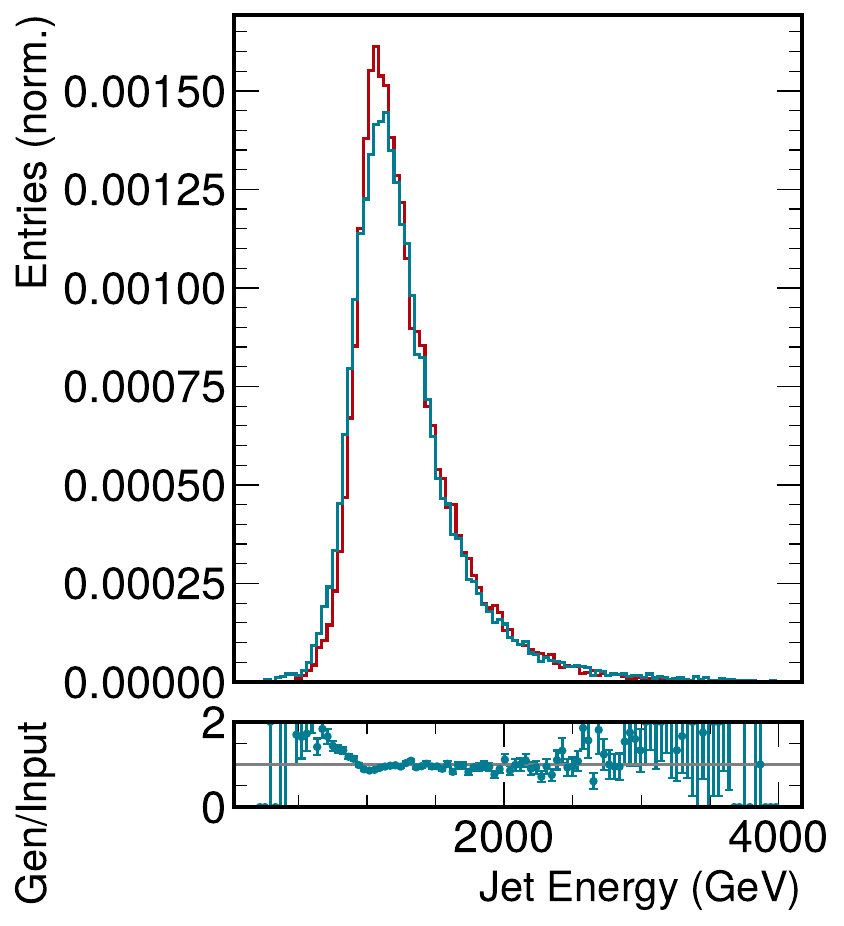}
    \includegraphics[width=0.3\textwidth]{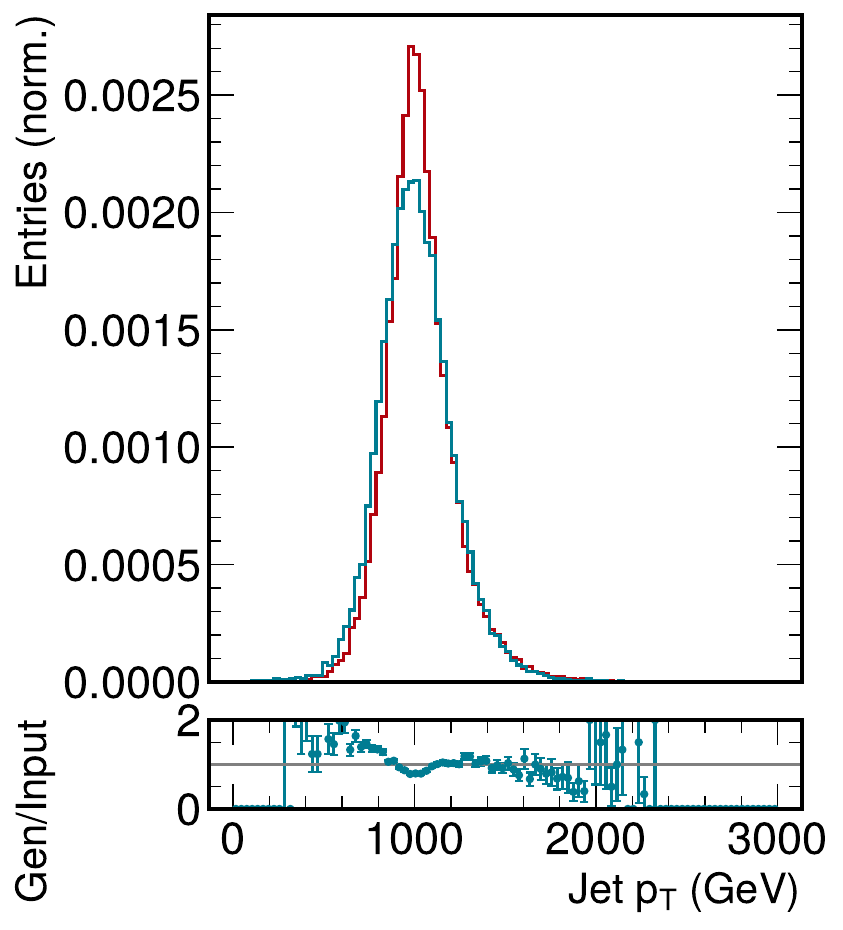} \\
    \includegraphics[width=0.3\textwidth]{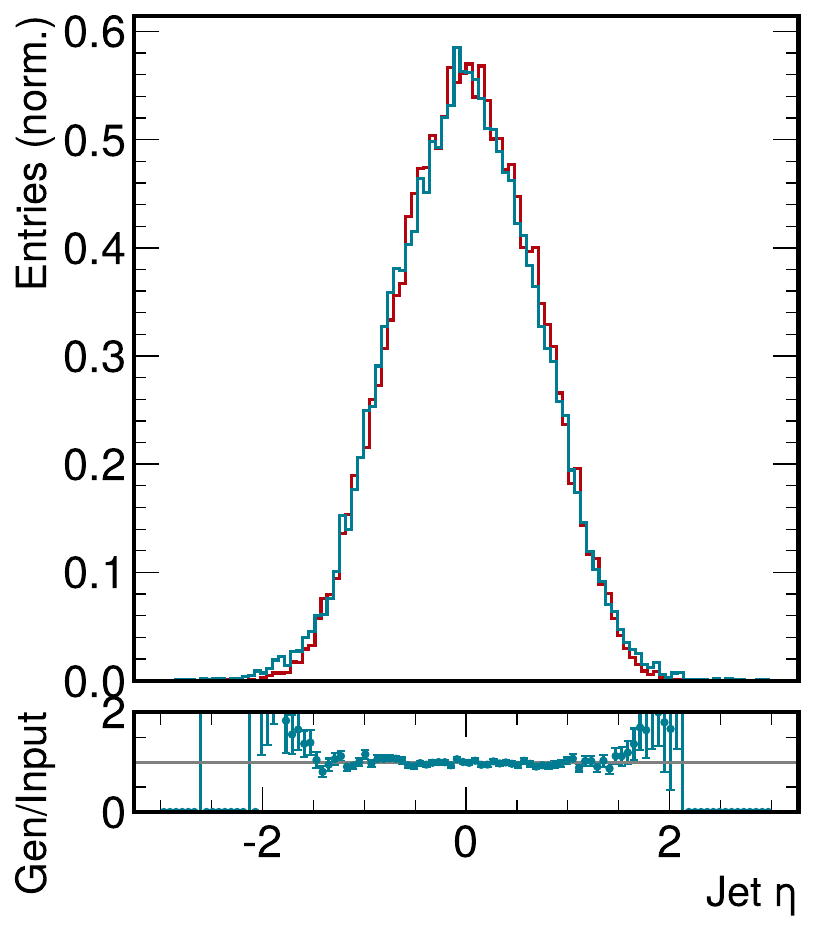}
    \includegraphics[width=0.3\textwidth]{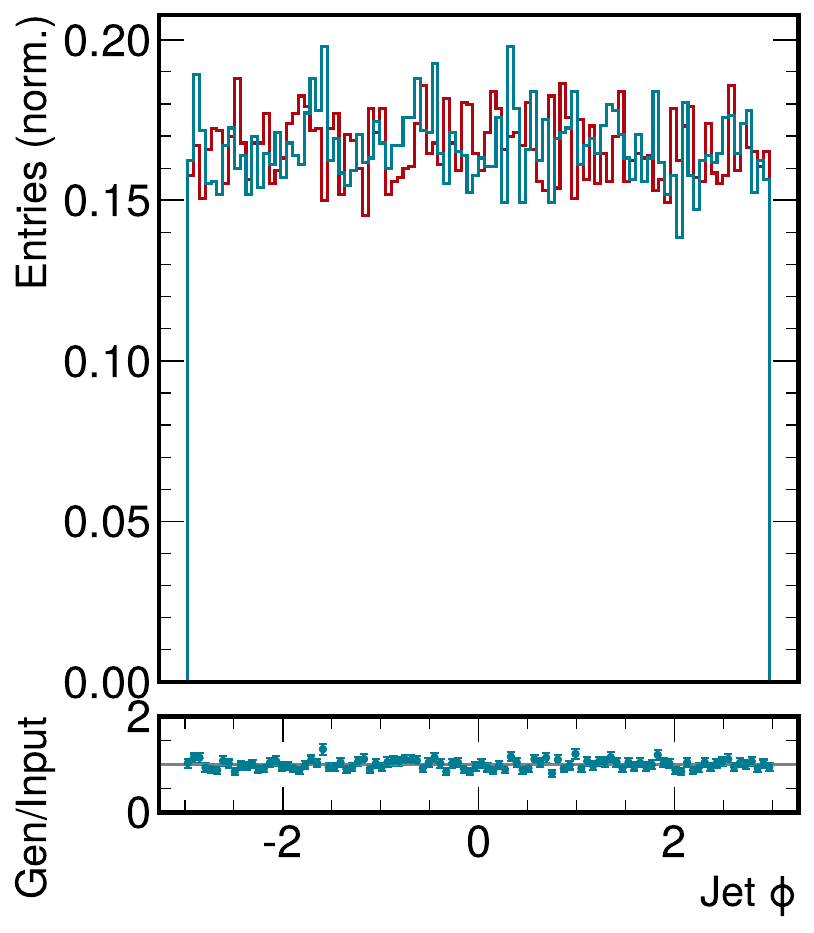}
    \caption{Comparison of input jets variable distributions (red) with randomly generated jets from the NF only model (blue). From left to right, top to bottom: mass, energy, \PT, $\eta$, and $\phi$ distributions are displayed. In the subpanels, the ratio $\sfrac{\mathrm{generated}}{\mathrm{input}}$ is shown.}
    \label{fig:NF}
\end{figure}

For comparison, a RealNVP network was also optimized, in the same way as the NF network for the ConVAE+NF approach, and trained to receive as input a flattened jet\footnote{The input jet 3$\times$30 feature matrix representing each particle's \px, \py and \pz, was flattened to a 1$\times$90 feature vector containing all particles' \px, then all \py and, finally, all \pz.}, transform each feature distribution into an uniform normal, and invert the transformation back to flattened jets. 
Table~\ref{tab:NF} contains the set of best parameters after hyperparameter optimization for this network.
The value of \emd for this case was 0.0049 and the histograms of jet variables are shown in Figure~\ref{fig:NF}.
The agreement between the distributions and the \emd are significantly worse than ConVAE+NF, demonstrating the advantage in combining the inference capability of the ConVAE, with the invertibility of the NF network.

\begin{table}[ht]
    \centering
    \begin{tabular}{c c c c}
    \toprule
         LR & $N_{\mathrm{flows}}$ & Layers & $N_\mathrm{linear}$  \\ \midrule
          3.4$\times 10^{-4}$ & 65 & 4 & [150, 200, 350, 150] \\
          \bottomrule
         %& 
    \end{tabular}
    \caption{Set of optimal hyperparameters for the NF.}
    \label{tab:NF}
\end{table}

Following the prior work in Refs.~\cite{MPGAN,Kansal:2022spb}, another set of metrics was calculated and compared to the results obtained using the state-of-the-art message-passing generative adversarial network (MPGAN) on this dataset.
The metrics $W_{1}^{M}$, $W_{1}^{P}$ and $W_{1}^\mathrm{EFP}$ correspond to the average 1-Wasserstein distance between input and generated jets mass, constituents $\eta^{\mathrm{rel}}$, $\phi^{\mathrm{rel}}$ and $\PT^{\mathrm{rel}}$, and five energy-flow polynomials (EFPs)~\cite{EFPs}, which are a set of variables that capture $n$-point particle correlations and jet substructure. 
The computation of the EFPs was performed using the \textsc{JetNet} package~\cite{JetNet}, where the EFPs correspond to the default choice defined in the package.
The Fr\'{e}chet ParticleNet Distance (FPND), which is the ParticleNet~\cite{ParticleNet}-adaptation of the Fr\'{e}chet Inception distance (FID)~\cite{FID} that is standard in computer vision; and coverage (COV) and minimum matching distance (MMD) which measure the ratio of samples in $X$ (e.g., the generated jets) that have a match in $Y$ (e.g., the input jets) and the average distance between matched samples, respectively, are also measured.
Table~\ref{tab:MPGAN} shows that among the models implemented in this work, the ConVAE+NF approach is the best in almost every metric. 
However, MPGAN outperforms the VAE-based approaches in every single metric.
The advantage of the ConVAE+NF approach is that it is approximately two times faster:
the average time to generate a single jet is $18.30 \pm 0.04$\unit{$\upmu$s} for ConVAE+NF versus 35.7\unit{$\upmu$s} for MPGAN measured using an NVIDIA Tesla T4 GPU.
% , given that its generated jets can be of use for some physics analysis applications, it can be useful in the future.

\begin{table}[ht]
    \centering
    \addtolength{\tabcolsep}{-0.1em}
    \begin{tabular}{c c c c c c c}
    \toprule
         Model & $W_{1}^{M}$ ($\times10^{-3}$) & $W_{1}^{P}$ ($\times10^{-3}$) & $W_{1}^\mathrm{EFP}$ ($\times10^{-5}$) & FPND & COV$\uparrow$ & MMD \\ \midrule
          ConVAE & 9.1 $\pm$ 0.6 & 8.5 $\pm$ 0.8 & 576 $\pm$ 1035 & 43.6 & 0.32 & 0.036 \\
          ConVAE+NF & \textbf{4.5 $\pm$ 0.5} & \textbf{5.3 $\pm$ 0.4} & \textbf{197 $\pm$ 247} & \textbf{34.3} & \textbf{0.38} & 0.034 \\
          NF & 12.7 $\pm$ 0.7 & 8.5 $\pm$ 0.8 & 8.6k $\pm$ 13.2k & 93.6 & \textbf{0.38} & \textbf{0.033} \\\midrule
          MPGAN & 0.7 $\pm$ 0.2 & 0.9 $\pm$ 0.3 & 0.7 $\pm$ 0.2 & 0.12 & 0.56 & 0.037 \\
          \bottomrule
         %& 
    \end{tabular}
    \caption{Set of metrics proposed in Refs.~\cite{MPGAN,Kansal:2022spb} to compare the performance of distinct ConVAE and NF approaches. In bold are the best values of the metrics when comparing the methods implemented in this work. The last row displays the performance of the MPGAN for the same gluon jet dataset.}
    \label{tab:MPGAN}
\end{table}

From the comparisons of the low- and high-level jet feature distributions, we observe that the mixed ConVAE+NF approach exhibits an improvement over the ConVAE alone.
We attribute this to two reasons: (1) the standard Gaussian distribution, used by the ConVAE alone but altered in the ConVAE+NF approach, is not the optimal probability distribution for the elements in the latent space dimension, and (2) the decompression of those values could be improved by using the same parameters of the ConVAE tuned for reconstruction.
Still, there are discrepancies between the ConVAE+NF generated jets and input jets, mainly related to the substructure variables, which are not present for the MPGAN approach.
Other than the structural distinction between GAN and VAE, one of the core differences between the two approaches is that the architecture for the MPGAN is a GNN.
The representation of jet constituents as an unordered set of particles is more natural, intrinsically preserves permutation invariance, and most likely contributes to the differences between the ConVAE and MPGAN results.

\section{Summary and Outlook}
\label{section:summary}

We presented a novel technique in the context of hadronic jets generation in simulated high-energy $pp$ collisions, based on a machine learning approach that combines a convolutional variational autoencoder (ConVAE) with normalizing flows (NFs).
From the values of \emd obtained, there was a clear improvement with respect to the previous work~\cite{VAESparse}, not only by the application of hyperparameter optimization to the ConVAE, but also with the usage of the ConVAE+NF technique.
The generation of gluon jets was performed using the ConVAE+NF model and compared with baseline ConVAE and NF models alone, as well as the state-of-the-art message-passing generative adversarial network (MPGAN).
The ConVAE+NF approach is superior to the first two, but needs further improvements to be comparable to the latter.

There is, however, an improvement in inference time compared to MPGAN as the ConVAE+NF can generate a jet twice as fast.
This can already offer advantages in physics applications where large amounts of simulated samples are needed but perfect accuracy is not a strict requirement.
It is worth noting that the MPGAN architecture is based on a graph neural network, which has a much more natural representation for a particle-based jet dataset. 
In future work, a graph-based VAE+NF can be implemented, trained, and tested to check for improvements in the jet generation capabilities.

\section*{Acknowledgements}

B.~O., J.~F., R.~C. and T.~T. are supported by grant 2018/25225-9, S\~{a}o Paulo Research Foundation (FAPESP).
B.~O. was also partially supported by grants 2019/16401-0 and 2020/06600-3, S\~{a}o Paulo Research Foundation (FAPESP).
R.~C. and T.~T were also partially supported by grant 2022/02950-5, S\~{a}o Paulo Research Foundation (FAPESP).
R.~K. was partially supported by an IRIS-HEP fellowship through the U.S. National Science Foundation (NSF) under Cooperative Agreement OAC-1836650.
R.~K. was additionally supported by the LHC Physics Center at Fermi National Accelerator Laboratory, managed and operated by Fermi Research Alliance, LLC under Contract No. DE-AC02-07CH11359 with the U.S. Department of Energy (DOE).
J.~D. is supported by the DOE, Office of Science, Office of High Energy Physics Early Career Research program under Award No. DE-SC0021187, the DOE, Office of Advanced Scientific Computing Research under Award No. DE-SC0021396 (FAIR4HEP), and the NSF HDR Institute for Accelerating AI Algorithms for Data Driven Discovery (A3D3) under Cooperative Agreement OAC-2117997.
M.~P. and M.~T. were supported by the European Research Council (ERC) under the European Union's Horizon 2020 research and innovation program (Grant Agreement No. 772369).
D.~G. is partially supported by the EU ICT-48 2020 project TAILOR (No. 952215).

\section*{References}

\bibliographystyle{lucas_unsrt}
\bibliography{bibliography.bib}

\end{document}